# Thermal and optical properties of two molecular potentials

Mahdi Eshghi [1], Sameer M. Ikhdair[2,3] , Ramazan Sever [4,*]

[1] *Department of Physics, Imam Hossein Comprehensive University, Tehran, Iran*

[2] *Department of Physics, Faculty of Science, An-Najah National University, Nablus, West Bank Palestine*

[3] *Department of Electrical Engineering, Near East University, Nicosia, Northern Cyprus, Mersin 10, Turkey*

[4,*] *Department of Physics, Middle East Technical University, Ankara, Turkey*

**Abstract**

We solve the Schrödinger wave equation for the generalized Morse and Cusp molecular potential models. In the limit of high temperature, at first, we need to calculate the canonical partition function which is basically used to study the behavior of the thermodynamic functions. Based on this, we further calculate the thermodynamic quantities such as the free energy, the entropy, the mean energy and the specific heat. Their behavior with the temperature has been investigated. In addition, the susceptibility for two level systems is also found by applying the incident time dependent field.

## 1. Introduction

When a particle is under the influence of a potential field, the thermodynamic interaction effects must be taken into consideration. Therefore, we need to find the exact analytical bound state solutions of the Schrödinger wave equation for a particle under the effect of these physical potentials. The potential models under present study are widely used in many branches of physics such as chemical physics, atomic physics, and nuclear physics (see for example [1-11]).

For example, the molecular Morse potential was studied widely since 1929 from various approaches. Therefore, many authors have solved the Schrodinger equation for some molecular potential models. A realization of the raising and lowering Ladder operators for the Morse potential was presented. Closed analytical expressions for the matrix elements of different operators that satisfy the commutation relations for the SU(2) group were found [12, 13]. The analytical solutions of the two-dimensional Schrodinger equation with the Morse potential were obtained by the series expansion method. The method was generalized to D-dimensional Schrodinger equation with Morse potential expanded in series about the origin. The series solutions of the Schrodinger equation with position-dependent mass for the Morse potential were found [14, 15]. The exact quantization rule method employed to obtain the arbitrary $l-$state energy solutions of the rotating Morse potential [16]. Further, the Morse potential in the momentum representation was presented analytically by hypergeometric functions [17, 18].

On the other hand, the approximate solution of the Dirac equation for the Rosen-Morse potential including the spin-orbit centrifugal term was obtained [19]. Overmore, at low momentum, the exact solutions of the bound and scattering states of the one-dimensional Dirac equation with an asymmetric cusp potential were studied [20].

The study of the thermodynamic functions for various potential models within non-relativistic regime has recently received much interest. For instance, many authors have used a method mainly based on the so called Huriwitz zeta function or even other different methods to analyze the thermodynamic properties of a few physical systems [21,22,23,24].

A recent study in this topic is wide. For example, the exact solutions of a one-dimensional Schrodinger equation with a harmonic oscillator plus inverse square potential were obtained to study the hidden symmetries and thermodynamic properties using the algebraic approach [25]. Further, a modified Rosen-Morse potential was solved for energy spectrum to study the thermodynamic properties by proper quantization rule [26].

In order to study the thermodynamic properties of the non-relativistic particles, we need to employ the canonical partition function which is usually considered as the building block in statistical thermodynamics as the role of the wave function in quantum physics. We first introduce the Schrödinger wave equation with Morse and Cusp molecular potential models. Followed by using some well-known appropriate methods, we can calculate the energy eigenvalues and the corresponding wave functions. Further, we attempt to analyze and find some of the thermodynamic quantities such as the free energy, the entropy, the mean energy and the specific heat for the Schrödinger wave equation with the generalized Morse and Cusp molecular potential models.

This work is organized as follows. In Sec. 2, we calculate the canonical partition function which is basically used to study the behavior of the thermodynamic quantities for the generalized Morse, and the Cusp potential models. Finally, in Section 3 we give our results and present the concluding remarks on these results.

## 2 The Schrödinger Wave Equation

The radial part of the time-independent Schrödinger equation can be written as

$$\frac{d^2 R_{n\ell}(z)}{dz^2} + \frac{2\mu}{\hbar^2}\left[E_n - V_{eff}(z)\right]R_{n\ell}(z) = 0, \tag{1}$$

with the effective potential that includes the rotational states has the form

$$V_{eff}(z) = V_\ell(z) + V(z), \quad V_\ell(z) = \ell(\ell+1)\hbar^2/2\mu z^2,$$

where $m$ is the mass of the particle and $E_{n\ell}$ is its non-relativistic binding energy.

### 2.1. The Generalized Morse Potential

This potential takes the general form [27, 28]

$$V(z) = V_1 \exp(-2\alpha z) - 2V_2 \exp(-\alpha z), \qquad z = (r - r_e)/r_e, \tag{2}$$

where $V_1$ and $V_2$ are the strength parameters of the potential.

In following Refs. [29, 30], the effective potential can simpler expanded in an exponential form as in (2) by using the Pekeris approximation:

$$V_{\text{eff}}(z) = \frac{\ell(\ell+1)\hbar^2}{2\mu r_e^2}\left[a_1 + a_2 \exp(-\alpha z) + a_3 \exp(-2\alpha z)\right]$$
$$+ V_1 \exp(-2\alpha z) - 2V_2 \exp(-\alpha z), \tag{3}$$

where the coefficients $a_i$, with $i = 1, 2, 3$ are defined as

$a_1 = 1 - 3/\alpha + 3/\alpha^2$, $a_2 = 4/\alpha + 6/\alpha^2$ and $a_3 = -1/\alpha + 3/\alpha^2$.

Now after substituting Eq. (3) into Eq. (1) along with some identifications:

$$A = -2\mu\left[\ell(\ell+1)a_3\hbar^2/2\mu r_e^2 + V_1\right]/\hbar^2,$$
$$B = -2\mu\left[\ell(\ell+1)a_2\hbar^2/2\mu r_e^2 - 2V_2\right]/\hbar^2, \tag{4}$$
$$C = -2\mu\left(E_{n\ell} - \ell(\ell+1)\hbar^2 a_1/2\mu r_e^2\right)/\hbar^2,$$

it becomes

$$\frac{d^2 R_{n\ell}(z)}{dz^2} - \left[Ae^{-2\alpha z} + Be^{-\alpha z} + C\right]R_{n\ell}(z) = 0, \tag{5}$$

After making further change of variables $y = \exp(-\alpha z)$ and replacing $R(z)$ by $R(y)$, Eq. (5) leads,

$$y^2 \frac{d^2 R_{n\ell}(y)}{dy^2} + y\frac{dR_{n\ell}(y)}{dy} - \frac{1}{\alpha^2}\left[Ay^2 + By + C\right]R_{n\ell}(y) = 0. \tag{6}$$

Now, defining a new wave function and a new variable,

$$\chi(y) = y^{-k} R(y), \qquad y = \exp(-\alpha z), \tag{7}$$

with $k$ to be calculated later on, then Eq. (6) can be rewritten in the form:

$$\alpha^2 y^2 \frac{d^2 \chi(y)}{dy^2} + (2k+1)\alpha^2 y \frac{d\chi(y)}{dy} + \left[(\alpha^2 k^2 - C) - (Ay^2 + By)\right]\chi(y) = 0. \tag{8}$$

If $k \in \Re^+$ which is specifically chosen to satisfy the condition:

$$\alpha^2 k^2 - C = 0 \to k = \sqrt{C}/|\alpha|, \tag{9}$$

then Eq. (8) can be transformed as

$$y\frac{d^2 \chi(y)}{dy^2} + (2k+1)\frac{d\chi(y)}{dy} - \frac{1}{\alpha^2}(Ay + B)\chi(y) = 0. \tag{10}$$

Since $A > 0$, we can pick up the following suitable ansatz for the wave function:

$$\chi(y) = \exp(\sqrt{A}\,y/\alpha) f(y), \tag{11}$$

then, Eq. (10) becomes a differential equation satisfying another function $f(y)$,

$$\alpha^2 y \frac{d^2 f(y)}{dy^2} + \left[(2k+1)\alpha^2 + 2\sqrt{A}\alpha y\right]\frac{df(y)}{dy} + \left[(2k+1)\alpha\sqrt{A} - B\right] f(y) = 0. \tag{12}$$

At this stage, we select $y = -\alpha\rho/2\sqrt{A}$, therefore we have

$$\rho \frac{d^2 f(\rho)}{d\rho^2} + [2k+1-\rho]\frac{df(\rho)}{d\rho} - \left[k + \frac{1}{2} - \frac{B}{2\alpha\sqrt{A}}\right] f(\rho) = 0,\ A > 0, \tag{13}$$

which is simply the known confluent hypergeometric equation [31, 32] with a given solution:

$$f(\rho) = {}_1F_1\left(k + \frac{1}{2} - \frac{B}{2\alpha\sqrt{A}}; 2k+1; \rho\right). \tag{14}$$

Now to calculate the wave function, we combine Eqs. (7), (9) and (11), we get the final result

$$R(y(z)) = y^{\frac{\sqrt{C}}{|\alpha|}} \exp\left(\frac{\sqrt{A}}{\alpha} y\right) {}_1F_1\left(k + \frac{1}{2} - \frac{B}{2\alpha\sqrt{A}}; 2k+1; \frac{2\sqrt{A}}{\alpha} y(z)\right), \tag{15}$$

where $y(z) = \exp(-\alpha z)$. Finally, we want to calculate the bound state energy equation. To achieve this, we follow Ref. [33] and find

$$k + \frac{1}{2} - \frac{B}{2\alpha\sqrt{A}} = -n. \tag{16}$$

Making use of Eqs. (4) and (9) together with Eq. (16), we can immediately arrive at the energy spectrum for any ro-vibrational $\ell$ state as

$$E_{n\ell} = D - \frac{\hbar^2 \alpha^2}{2\mu r_e^2}\left[\left(n + \frac{r_e\sqrt{2\mu}}{2\hbar\alpha}\left(\frac{\left[2V_2 - \frac{\ell(\ell+1)\hbar^2}{2\mu r_e^2}\left(\frac{4}{\alpha} + \frac{6}{\alpha^2}\right)\right]}{\sqrt{V_1 + \frac{\ell(\ell+1)\hbar^2}{2\mu r_e^2}\left(-\frac{1}{\alpha} + \frac{3}{\alpha^2}\right)}}\right)\right) + \frac{1}{2}\right]^2, \tag{17}$$

where $D = \frac{\hbar^2}{2\mu r_e^2} \ell(\ell+1)\left(1 - \frac{3}{\alpha} + \frac{3}{\alpha^2}\right)$. Our results are in high agreement with Refs. [33, 34]. Further, for the vibrational states $\ell = 0$, we can obtain the energy spectrum of the molecular Morse potential as

$$E_n = -\frac{2\alpha^2 \hbar^2}{8\mu r_e^2}\left(2n + 1 - \frac{r_e\sqrt{2\mu V_2}}{\hbar \alpha \sqrt{V_1}}\right)^2. \tag{18}$$

As a result, it is obvious that the energy spacing difference between levels decreases by decreasing the well width.

*a) Thermal properties:*

To study the thermal properties, we can follow Ref. [35]. Therefore, as a first step, we seek to find the canonical partition function $Z$ and study the thermodynamic properties of the molecular Morse potential model. At first, we recast Eq. (17) in a rather simpler form as

$$E_{n\ell} = D - \frac{\hbar^2 \alpha^2}{2\mu r_e^2}\left[\frac{1}{4}(4n - \Xi)\right]^2, \tag{19}$$

where $\Xi = -\frac{2r_e\sqrt{2\mu}}{\hbar \alpha}\left[\left[2V_1 - \frac{\ell(\ell+1)\hbar^2}{2\mu r_e^2}\left(\frac{4}{\alpha} + \frac{6}{\alpha^2}\right)\right] \Big/ \sqrt{\left[V_1 + \frac{\ell(\ell+1)\hbar^2}{2\mu r_e^2}\left(-\frac{1}{\alpha} + \frac{3}{\alpha^2}\right)\right]}\right] - 2$,

$n = 0,1,2,...,n_{max} \leq \left|\frac{\Xi}{4}\right|$. It is worth noting that in the Morse potential, the quantum number $n$ cannot be taken as infinite large number, there exists an upper limit $n_{max}$. This is because the bound states of the Morse potential are finite [36].

In addition, we introduce the canonical partition function in the form:

$$Z(\beta, \Xi) = \sum_{n=0}^{\Xi} \exp\left(-\frac{E_{n\ell}}{k_B T}\right), \tag{20}$$

where $T$ and $k_B$ are the temperature and the Boltzmann constant, respectively. On substituting Eq. (19) into Eq. (20), the canonical partition function, at temperature $T$, can be easily calculated through the Boltzmann factor as

$$Z(\beta,\Xi) = e^{-D} \sum_{n=0}^{\Xi} \exp\left(\left[\frac{1}{4\sigma}(4n-\Xi)\right]^2\right), \tag{21}$$

with the following identifications

$$\beta = (k_B T)^{-1}, \quad \tau = \frac{\sqrt{2\mu r_e}}{\hbar\alpha}, \quad \sigma = \tau\beta^{-1/2}. \tag{22}$$

At high temperature, we can use the approximation $\exp\left[-\frac{\hbar^2}{2\mu r_e^2}\ell(\ell+1)\left(1-\frac{3}{\alpha}+\frac{3}{\alpha^2}\right)\right] \approx 1$, Therefore, the partition function $Z$ becomes

$$Z(\beta,\Xi) = \sum_{n=0}^{\Xi} \exp\left(\left[\frac{1}{4\sigma}(4n-\Xi)\right]^2\right). \tag{23}$$

Under the classical limit, while considering the above approximation and with large $\Xi$ and small $\beta$, the sum can be replaced by the integral form:

$$Z(\beta,\Xi) = \sigma \int_0^{\frac{\Xi}{4\tau}\sqrt{\beta}} e^{t^2} dt = -\frac{\tau\sqrt{\pi}\, Erfi\left(\frac{\Xi}{4\tau}\sqrt{\beta}\right)}{2\sqrt{\beta}}, \quad y=(\Xi-4n)/4\sigma. \tag{24}$$

Moreover, it is well known that the imaginary error function in Eq. (24) is called an entire function and can be defined as $Erfi(x) = iErf(ix)$. Here, $Erf$ denotes the error function which is a special function of sigmoid shape that occurs in probability, statistics and partial differential equations.

In maple and mathematica software codes, the error function can be defined as $\frac{2}{\sqrt{\pi}}\int_0^x e^{t^2} dt = Erfi(x)$ [22, 32]. This imaginary error function can be implemented in mathematica as $Erfi[x]$.

The other thermodynamic properties of the system can be easily obtained from partition function. In fact, any other parameter that might contribute to the energy should also appear in the argument of $Z$ [35]. Here we display these thermal quantities as follows. The mean energy $U$ can be defined as

$$U(\beta,\Xi) = -\frac{\partial}{\partial \beta}\ln Z(\beta,\Xi) =$$

$$= -\frac{k_B T \frac{\Xi}{4\tau}}{\sqrt{\pi}\, Erfi\!\left(\frac{\Xi}{4\tau\sqrt{k_B T}}\right)} \left[\exp\!\left(\frac{\Xi}{4\tau\sqrt{k_B T}}\right)^{\!2} - \frac{\sqrt{\pi}\, Erfi\!\left(\frac{\Xi}{4\tau\sqrt{k_B T}}\right)}{\frac{\Xi}{2\tau}}\right]. \quad (25)$$

Figure 1 plots the variation of the mean energy $U$ (in $MeV$ units) with the temperature $T$ for a nonrelativistic particle in the field of the generalized Morse potential. For the rotational $\ell = 2$ state, it is obvious that $U$ increases with the increasing temperature $T$ until it reaches its maximum value at large temperature at nearly $100\, K$.

The $Erfi(x)$ function in Eq. (25) is appeared as [22, 32]:

$$e^{-x^2}\int_0^x e^{t^2}\, dt = \frac{\sqrt{\pi}}{2} e^{-x^2} erfi(x) = dowson(x).$$

Further, the Dowson's integral is defined in mathematica as $Dowson\, F[x]$. Also, Dawson's integral is defined in maple as follows: $dawson(x) = e^{-x^2}\left(\int_0^x e^{t^2}\, dt\right)$.

Next, the specific heat $C$ is being calculated as

$$C(\beta,\Xi) = -\frac{\partial}{\partial T} U(\beta,\Xi)$$

$$= k_B\left(\frac{1}{4\sqrt{T}} - T\right) + \frac{1}{4T} + \frac{\Xi \exp\!\left(\frac{\Xi}{4\tau\sqrt{k_B T}}\right)^{\!2}}{\tau\sqrt{\pi}\, Erfi\!\left(\frac{\Xi}{4\tau\sqrt{k_B T}}\right)}\left[\frac{1}{\sqrt{k_B T}}\left(\frac{k_B}{4} + \frac{9}{8T} + \left(\frac{\Xi}{8\tau}\right)^{\!2}\right)\right]$$

$$+ \frac{\Xi k_B \exp\!\left(\frac{\Xi}{4\tau\sqrt{k_B T}}\right)}{4\tau\sqrt{\pi}(k_B T)^3 Erfi\!\left(\frac{\Xi}{4\tau\sqrt{k_B T}}\right)^{\!2}}\left[\frac{1}{\sqrt{k_B T}}\left(Erfi\!\left(\frac{\Xi}{4\tau\sqrt{k_B T}}\right) + \frac{\exp\!\left(\frac{\Xi}{4\tau\sqrt{k_B T}}\right)^{\!2}\Xi}{2\tau\sqrt{\pi k_B T}}\right)\right]. \quad (26)$$

Figure 2 shows the behavior of the specific heat $C$ versus $T$. In this regards, it is seen that the specific heat decreases with the increasing temperature. Notice that the specific heat of a nonrelativistic particle under the influence of the generalized Morse potential approaches zero for the rotational $\ell = 2$ state.

Now, the Helmholtz free energy $F$ takes the form

$$F(\beta, \Xi) = -k_B T \ln\left(\frac{\tau\sqrt{\pi k_B T}}{2} Erfi\left(\frac{\Xi}{4\tau\sqrt{k_B T}}\right)\right). \tag{27}$$

Fig. 3a shows the behavior of the free energy $F$ with the temperature $T$. At very low temperatures nearly $0\,C \leq T \leq 2\,C$ the free energy goes down in the negative region (strongly attractive) then increases (weakly attractive) and starts to increase in the positive region (repulsive). In On the other hand, in Fig. 3b, the free energy increases rapidly in the positive region and becomes strongly attractive at high temperatures.

Now, here we write a description for the reminders about the entropy then calculate the entropy from the partition function. Finally, the entropy is a fundamental measure of information content which can be applied in a wide variety of fields. Further, it plays an important role in thermodynamics and is a central second law of thermodynamics. Entropy helps to measure the amount of order and disorder and/or chaos and can be defined and measured in many other fields than the thermodynamics. For instance, in classical physics, entropy is defined as the quantity of energy incapable of physical movements. Thus, the entropy $S$ given as

$$\begin{aligned}S(\beta, \Xi) &= k_B \ln Z(T, \Xi) + k_B T^2 \frac{\partial}{\partial T} \ln Z(T, \Xi) \\ &= 2k_B \ln\left(\frac{\tau}{2}\sqrt{k_B T}\sqrt{\pi}.Erfi\left(\frac{\Xi}{4\tau\sqrt{k_B T}}\right)\right) \\ &\quad - \frac{2k_B}{\tau\sqrt{\pi}Erfi\left(\frac{4\Xi}{\sqrt{k_B T}}\right)}\left(\frac{2\tau T \Xi \exp\left(-\frac{16\Xi^2}{k_B T}\right)}{\sqrt{k_B T}} - \tau T\sqrt{\pi}Erfi\left(\frac{4\Xi}{\sqrt{k_B T}}\right)\right).\end{aligned} \tag{28}$$

Fig. 4a shows the behavior of the entropy $S$ at low temperature $T$. The entropy falls down in the interval nearly $0\,C \leq T \leq 2\,C$ but remains in the positive (repulsive) regime then starts to increase back and returns to its initial value in the interval nearly $2\,C \leq T \leq 10\,C$.

However, in Fig. 4b, the entropy increases rapidly in the positive regime with high temperature.

We can also extend our calculations of the canonical partition function to an interaction-free $N$-body system via $Z = Z^N$. The dependence on $N$ and volume comes via the dependence on the energy eigenvalues $E_{n\ell}$.

b) *Optical properties*

Now, let us consider a two-level system with an electric field applied in $z$-direction of this potential as $\vec{F} = e|\vec{E}|z$. The Hamiltonian of the system splits into two parts with the interaction of an electromagnetic wave in the dipole approximation is: $H = H_0 + e|\vec{E}|z$, where $H_0$ and $|\vec{E}|$ are the unperturbed part and the external electric field, respectively [23, 24], and $e$ is the absolute value of the electron charge.

Using the time-independent, non-degenerate perturbation theory [37], if $e|\vec{E}|z << |E_0^{(0)} - E_1^{(0)}|$ is satisfied, we can obtain the corresponding energy levels as follows

$$\Omega_n = E_n - E_n^{(0)} = e|\vec{E}|z + e^2|\vec{E}|^2 \sum_{j \neq n} \frac{|z_{nj}|^2}{E_n^{(0)} - E_j^{(0)}} + ..., \qquad (29)$$

and the wave function as follows

$$\phi_n(z) = \phi_n^{(0)}(z) + e|\vec{E}| \sum_{j \neq n} \phi_j^{(0)}(z) \frac{|z_{nj}|}{E_n^{(0)} - E_j^{(0)}} + ..., \qquad (30)$$

where no degeneracy for this state, $\phi_0^{(0)}$ is expected to be a parity eigenstate. Hence, $z_{nn} = 0$ and by using the unperturbed wave functions, the matrix elements are evaluated.

To obtain of the first-order and third-order susceptibilities, first, we apply the incident field as follows: $E(t) = \sum_j E(\omega_j) \exp(-i\omega_j t)$. Now, by using the compact density-matrix approach [38] and based on the Fermi golden rule in quantum mechanic, we obtain the linear susceptibility, namely, first-order susceptibility as

$$\eta^{(1)}(\omega) = \frac{\rho_s |M_{10}|^2}{E_{10} - \hbar\omega_{ij} - i\hbar\Gamma_0}, \qquad (31)$$

and third order nonlinear susceptibility as

$$\eta^{(3)}(\omega, I) = \frac{2\pi I \rho_s |M_{10}|^4}{n_r c (E_{10} - \hbar\omega_{ij} - i\hbar\Gamma_0)} \left[ \frac{4}{(E_{10} - \hbar\omega_{ij})^2 + (\hbar\Gamma_0)^2} \right.$$
$$\left. - \frac{|M_{11} - M_{00}|^2}{|M_{10}|^2} \frac{1}{(E_{10} - i\hbar\Gamma_0)(E_{10} - \hbar\omega_{ij} - i\hbar\Gamma_0)} \right], \qquad (32)$$

where $E_{ij} = E_i - E_j$ and $M_{ij} = \langle \phi_i | qz | \phi_j \rangle \delta_{k_n k'_n}$, and $\rho_s$, $n_r$ and $I$ are the density, refractive index and incident optical intensity, respectively. Therefore, we can write the total susceptibility as

$$\eta(\omega, I) = \eta^{(1)}(\omega) + \eta^{(3)}(\omega, I). \qquad (33)$$

## 2.2. Cusp potential

For several reasons, a Cusp potential is of much interest in physics. If the parameter $V_0$ becomes attractive $V_0 < 0$, it can be regarded as a screened one-dimensional Coulomb potential [39]. In addition, for theoretical prediction of many physical properties of diatomic molecules requires knowledge of the radial wave functions of its scattering states and phase shifts. Furthermore, there is one noticeable difference between the Wood-Saxon and Cusp potentials. The Cusp potential does not exhibit a square barrier limit.

Now, if we take in Eq. (2), $V_1 = 0$, $-2V_2 = V_0$ and $\alpha = 1/\tilde{A}$, then the generalized Morse potential turn out to the Cusp potential. This Cusp potential has the form:

$$V(x) = V_0 \exp(-\alpha|z|), \quad \alpha = 1/\tilde{A}, \qquad (34)$$

where the parameter $V_0$ shows the height of barrier or the depth of the well and the positive constant $\tilde{A}$ determines the shape of the potential [40. Substituting this potential into Eq. (1) for $z > 0$, we obtain

$$\frac{d^2 R_{n\ell}(z)}{dz^2} + \frac{2\mu}{\hbar^2}\left[ E_n + V_0 \exp(-\alpha z) - \frac{\ell(\ell+1)\hbar^2}{2\mu z^2} \right] R_{n\ell}(z) = 0. \tag{35}$$

To avoid any repetition in the above calculations, we can only obtain the energy and the wave function for this potential with the substitution of $V_1 = 0$, $-2V_2 = V_0$ and $\alpha = 1/\tilde{A}$ into Eq. (17) and (15), and other calculated thermodynamic and optical quantities, too.

## 3. Results and Concluding Remarks

In this work, we studied the thermodynamic quantities of a nonrelativistic particle moving under the field of the generalized Morse and the Cusp potentials. We have also investigated the optical properties for two-level systems for the generalized Morse potential. The thermal quantities like the mean energy increases with the increasing temperature until it becomes static at high temperature, However the specific heat decrease with the increasing temperature and approaches to zero at high temperature.

The behavior of the free energy with temperature has been shown. At very low temperature it is like an attractive well. However, it increases rapidly at high temperature (the repulsive regime). Finally the entropy is in the repulsive well at relatively low temperature but increase rapidly at high temperature.

At the end, it is found that the cusp potential has a very similar behavior as the generalized Morse potential.

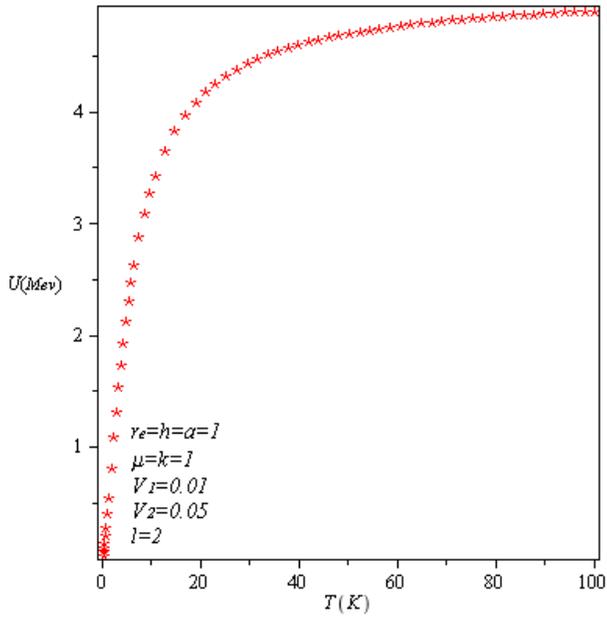

**Fig. 1** The behavior of the mean energy $U$ with the temperature $T$

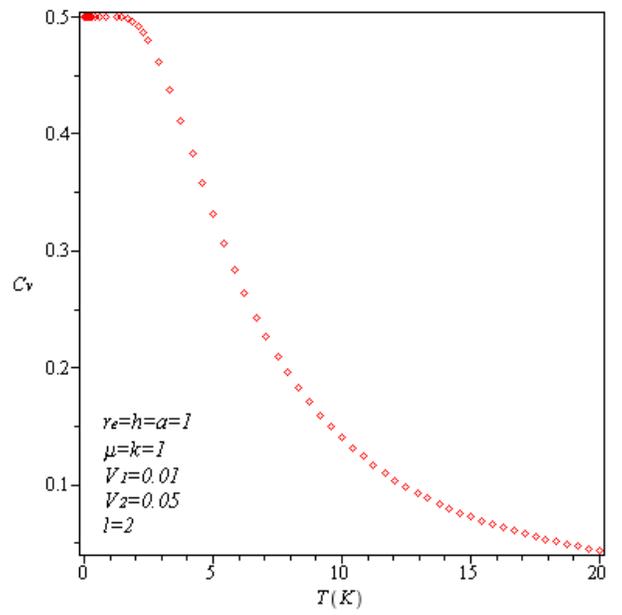

**Fig. 2** The behavior of the specific heat $C$ with the temperature $T$.

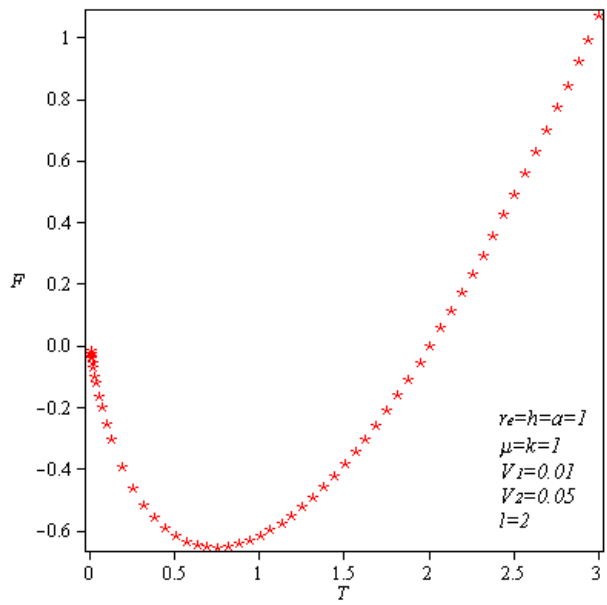

**Fig 3a.** The behavior of the free energy $F$ with the low temperature $T$

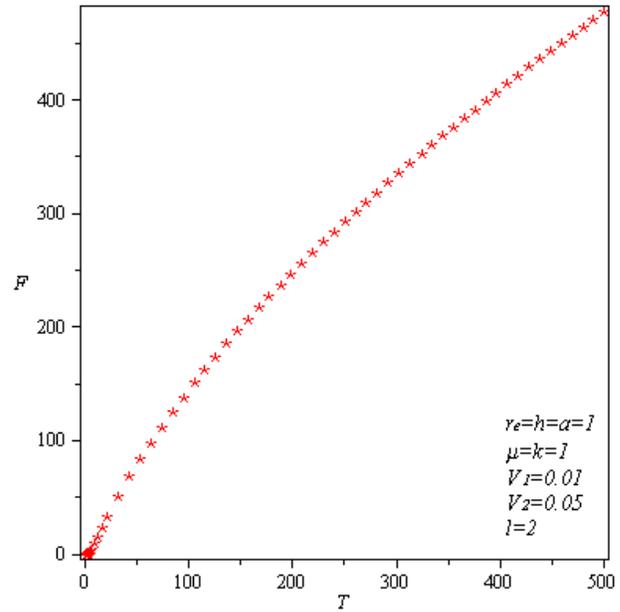

**Fig 3b.** The behavior of the free energy $F$ with the high temperature $T$

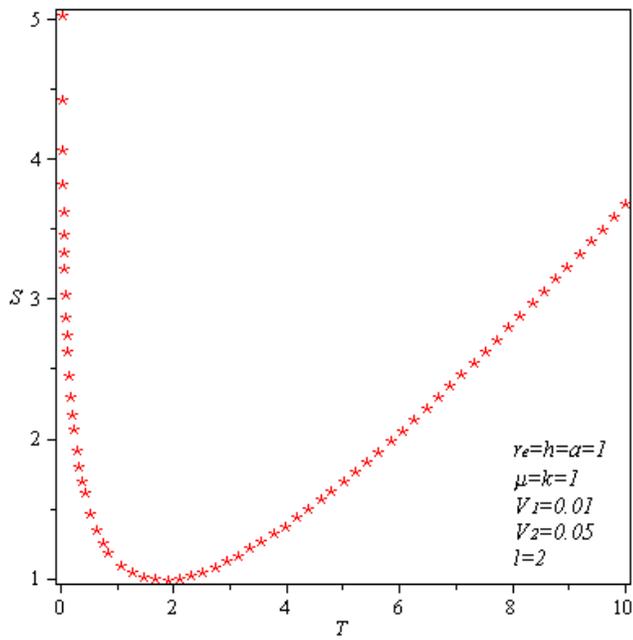 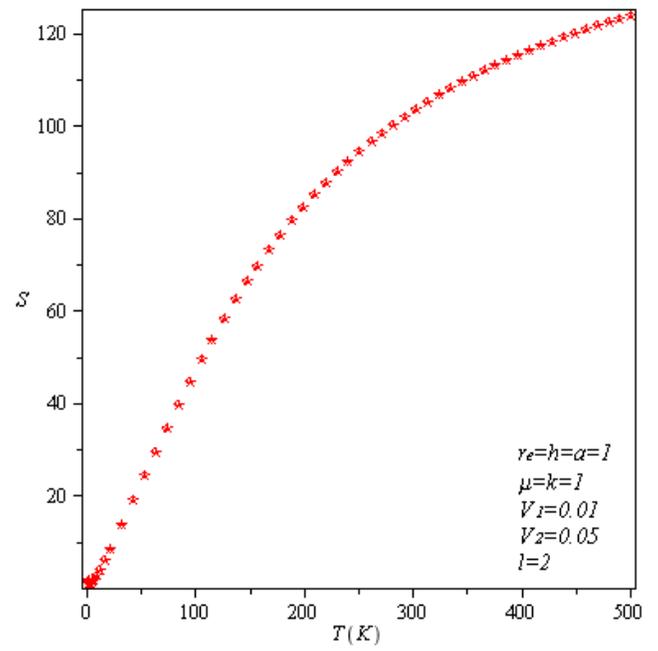

**Fig 4a.** The behavior of the entropy $S$ with the low temperature $T$

**Fig 4b.** The behavior of the entropy $S$ with the high temperature $T$